\documentclass{aa}
\usepackage{graphicx}
\usepackage{txfonts}
\usepackage{hyperref}
\begin{document}

   \title{Organic matter in interstellar dust lost at the approach to the heliosphere}

   \subtitle{Exothermic chemical reactions of free radicals ignited by the Sun}

   \author{Hiroshi Kimura
          \inst{1}
          \and
          Frank Postberg\inst{2,3}
          \and
          Nicolas Altobelli\inst{4}
          \and
          Mario Trieloff\inst{3,5}
          }

   \institute{Planetary Exploration Research Center (PERC), Chiba Institute of Technology, 
Tsudanuma 2-17-1, Narashino, Chiba 275-0016, Japan\\
              \email{hiroshi\_kimura@perc.it-chiba.ac.jp}
         \and
            Institut f\"{u}r Geologische Wissenschaften, Freie Universit\"{a}t Berlin, Malteserstr. 74-100, 12249 Berlin, Germany
         \and
             Institute of Earth Sciences, University of Heidelberg, Im Neuenheimer Feld 234-236, D-69120 Heidelberg, Germany
         \and
             Solar System Science Operation Division, Directorate of Science and Robotic Exploration, \\
    ESA-ESAC - P.O. Box 78, E-28691 Villanueva de la Ca\~{n}ada, Madrid, Spain
         \and
             Klaus Tschira Laboratory for Cosmochemistry, University of Heidelberg, Im Neuenheimer Feld 234-236, D-69120 Heidelberg, Germany
             }

   \date{Received July 16, 2015; accepted September 15, 2020}

% \abstract{}{}{}{}{} 
% 5 {} token are mandatory
 
  \abstract
   {}
   {We tackle the conundrums of organic materials missing from interstellar dust when measured inside the Solar System, while undoubtedly existing in the local interstellar cloud (LIC), which surrounds the Solar System.}
   {We present a theoretical argument that organic compounds sublimate almost instantaneously by exothermic reactions, when solar insolation triggers the recombination of free radicals or the rearrangement of carbon bonds in the compounds.}
   {It turns out that the triggering temperature lies in the range of 20--50~K by considering that sublimation of organic materials takes place beyond the so-called filtration region of interstellar neutral atoms.
We find that in-situ measurements of LIC dust in the Solar System result in an overestimate for the gas-to-dust mass ratio of the LIC, unless the sublimation of organic materials is taken into account.
We also find that previous measurements of interstellar pickup ions have determined the total elemental abundances of gas and organic materials, instead of interstellar gas alone.}
   {We conclude that LIC organic matter suffers from sublimation en route to the heliosphere, implying that our understanding of LIC dust from space missions is incomplete. 
   Since space missions inside the orbit of Saturn cannot give any information on the organic substances of LIC dust, one must await a future exploration mission to the inner edge of the Oort cloud for a thorough understanding of organic substances in the LIC.
   Once our model for the sublimation of interstellar organic matter by exothermic chemical reactions of free radicals is confirmed, the hypothesis of panspermia from the diffuse interstellar medium is ruled out.}

   \keywords{dust, extinction --
                ISM: abundances --
                ISM: clouds --
                ISM: individual objects: the local interstellar cloud (LIC)
               }
   \maketitle
%
%________________________________________________________________

\section{Introduction}

Our Solar System is currently immersed in a warm, partially ionized, diffuse interstellar medium (DISM), called the local interstellar cloud (LIC), consisting of gas and dust.
The motion of the Sun relative to the LIC results in a unidirectional wind of the LIC materials toward the Sun and the formation of the heliospheric boundary due to the interaction between the interstellar wind and the solar wind \citep{bertaux-blamont1971,frisch-et-al1999}.
While interstellar neutral atoms and large dust particles penetrate the heliospheric boundary and are detected inside the heliosphere, interstellar ions and tiny dust particles are filtered out at the boundary \citep{kimura-mann1998,kimura-mann1999,linde-gombosi2000,czechowski-mann2003}.
In the 1970s, possible detections of LIC dust were reported by \citet{bertaux-blamont1976} through an analysis of impact data from capacitor-type detectors onboard the Meteoroid Technology Satellite and by \citet{wolf-et-al1976} from the multi-coincidence microparticle sensing system onboard Pioneer 8 and 9 \citep[see also][]{mcdonnell-berg1975}.
In the late part of the 20th century, the stream of dust particles from the LIC was unambiguously recorded by impact ionization dust detectors onboard Ulysses, Galileo, Hiten, Nozomi, and Cassini \citep{gruen-et-al1994,gruen-et-al1997,svedhem-et-al1996,sasaki-et-al2007,altobelli-et-a2003}.
For a thorough review of LIC dust measurements performed during the 20th century, see \citet{mann-kimura2000}.

In the last decade, our understanding of LIC dust has been greatly improved by data analyses of recent space missions, data mining of previous space missions, elaborate numerical simulations of dust dynamics, and comprehensive studies of gas depletion measurements. 
An analysis of Helios in-situ dust data have revealed that the time-of-flight mass spectra of interstellar dust are dominated by silicates and iron \citep{altobelli-et-a2006}.
NASA's Stardust mission was successful in capturing a collection of LIC dust particles and bringing them back to Earth for a thorough analysis in a laboratory \citep{frank-et-al2014}.
The Stardust samples of LIC dust are mineral grains, but they show no clear evidence for the presence of organic refractory material in the LIC \citep{bechtel-et-al2014,westphal-et-al2014b}.
The absence of organic refractory materials in a near-Earth orbit is not a complete surprise because the organic-rich population of LIC dust may not be able to easily reach the inner Solar System against high radiation pressure repulsion from the Sun \citep{landgraf1999,kimura-et-al2003b,sterken-et-al2012}.
The impact ionization time-of-flight mass spectrometer onboard Cassini recorded signals of 36 LIC dust grains in the proximity of Saturn and revealed that LIC dust is mainly composed of Mg-rich silicates; some of them are mixed with Fe-bearing metals and/or oxides, but not with organic compounds \citep{altobelli-et-al2016}.
Since the penetration of organic-rich carbonaceous grains could reach the orbit of Saturn, they should have been detected by Cassini if they exist.
Nevertheless, we cannot exclude the possibility that organic-rich carbonaceous grains are simply smaller than the detection threshold of the Cassini dust analyzer (i.e., grain radius of $a \sim 20~\mathrm{nm}$).
The most recent model of interstellar dust by \citet{jones-et-al2013} suggests that carbonaceous grains in the DISM are shattered into tiny grains of $a < 20~\mathrm{nm}$ by interstellar shocks.
However, the heavy depletion of organic materials in submicrometer-sized grains looks incongruous since there is no trace of such a strong shock in the LIC \citep{kimura2015}.

The gas-depletion measurements of the LIC by the Hubble Space Telescope (HST) indicate that LIC dust contains refractory organic material consisting of C, H, O, and N \citep{kimura-et-al2003a,kimura-et-al2003b,kimura2015}.
There is, therefore, a remarkable discrepancy that a significant organic-rich population of interstellar dust exists in the LIC, but it has never been identified inside the Solar System.
Another conundrum emerges from the composition of gas in the LIC determined by in-situ measurements of interstellar pickup ions in the inner Solar System.
The chemical composition of the LIC derived from pickup-ion measurements indicates that no N atoms are incorporated into LIC dust \citep{gloeckler-geiss2004}.
This contradicts the fact that the HST measurements of gas absorption lines along the line of sight toward nearby stars have revealed the depletion of N atoms in the gas phase of the LIC \citep{wood-et-al2002,kimura-et-al2003a}.
Moreover, nearly half of the O atoms are depleted in the gas phase of the LIC, contrary to the pickup-ion measurements indicating that the majority of the O atoms reside in the gas phase \citep{gloeckler-geiss2004,kimura2015}.
As a result, the discrepancy between the pickup ion measurements and the gas absorption measurements remains a deep mystery.

The purpose of this study is to solve the conundrums of the missing organic materials in interstellar dust streaming into the Solar System from the LIC.
The most reasonable hypothesis would be the loss of organic refractory materials from interstellar dust en route to the inner Solar System.
Here, we show that all the measurements of the LIC materials are in harmony if the sublimation of organic compounds from LIC dust proceeds from exothermic reactions in the organic substances.

%__________________________________________________________________

\section{Model}

We propose that the missing organics problems can be solved if the elements forming organic compounds (i.e., C, H, O, and N) desorb from LIC dust near the Sun.
There are several energetic processes that could help C, H, O, and N to desorb from organic refractory material in the interstellar medium \citep{baragiola-et-al2005,collings-mccoustra2012}.
Among these processes, we consider that exothermic reactions would release sufficient energy to sublimate organic materials when interstellar dust is heated by solar radiation to a temperature that is high enough to trigger the reactions.
Exothermic reactions are associated with a spontaneous release of energy by the recombination of reactive atoms and molecules in the organic materials.
Laboratory experiments on the recombination of free radicals or the rearrangement of carbon bonds have revealed that such exothermic reactions are accompanied by explosive events \citep{dhendecourt-et-al1982,benit-roessler1993,wakabayashi-et-al2004}.

As a typical picture of LIC dust, hereafter, we consider a particle that has a radius of $a=0.1~\mathrm{\mu m}$ and consists of amorphous silicates in the core of the particle and organic material in the mantle \citep{li-greenberg1997}.
The mass fraction $x$ of organic material in LIC dust is $x = 0.39$, based on the most plausible assignment of dust-forming elements to the composition of dust in the LIC (see Appendix~\ref{appendix_a}).
The mass of the particle, $m_\mathrm{p}$, is given by $m_\mathrm{p} = \left({4 / 3}\right) \pi a^{3} \left[{x / \rho_\mathrm{or} +\left({1-x}\right) / \rho_\mathrm{sil}}\right]^{-1}$ where $\rho_\mathrm{sil}$ is the density of the silicate core and $\rho_\mathrm{or}$ is the density of the organic mantle.
By assuming $\rho_\mathrm{sil} = 3.5 \times {10}^{3}~\mathrm{kg~m^{-3}}$ and $\rho_\mathrm{or} = 1.8 \times {10}^{3}~\mathrm{kg~m^{-3}}$, we obtain $m_\mathrm{p} \simeq 1.1\times{10}^{-17}~\mathrm{kg}$ for the core-mantle particles with $a=0.1~\mathrm{\mu m}$ \citep[cf.][]{li-greenberg1997}.

The energy, $\varepsilon$, required to raise a dust particle to the temperature, $T$, from the triggering temperature of exothermic chain reactions, $T_\mathrm{trig}$, is given by 
\begin{eqnarray}
\varepsilon\left({T}\right) = m_\mathrm{p} \int_{T_\mathrm{trig}}^{T} C_\mathrm{p}\left({T'}\right) \,dT', \label{required-energy}
\end{eqnarray}
where $C_\mathrm{p}\left({T}\right)$ is the specific heat of the particle at the temperature $T$ \citep{leger-et-al1985,sorrell2001,duley-williams2011}.
The specific heat at $T=298~\mathrm{K}$ lies in the range of $C_\mathrm{or} = 0.7$--$1.9~\mathrm{kJ~kg^{-1}~K^{-1}}$ for organic materials and $C_\mathrm{sil} = 0.74$--$0.86~\mathrm{kJ~kg^{-1}~K^{-1}}$ for silicate materials \citep{domalski-hearing1990,campbell-norman1998,winter-saari1969,zeller-pohl1971,krishnaiah-et-al2004}.
Accordingly, we estimate the specific heat of a silicate-core, organic-mantle particle to be $C_\mathrm{p} = 1.0~\mathrm{kJ~kg^{-1}~K^{-1}}$ at $T=298~\mathrm{K}$ using $C_\mathrm{p} = x\,C_\mathrm{or} + \left({1-x}\right) C_\mathrm{sil}$ \citep{senshu-et-al2002}.
For the sake of simplicity, we may approximate the dependence of specific heat on the temperature by $C_\mathrm{p}\left({T}\right) \propto T$ in the temperature range of interest \citep[cf.][]{sharp-ginther1951,wong-westrum1971,kay-goit1975,richet-et-al1982}.

We would like to point out that desorption of organic forming elements only takes place if $\varepsilon$ is less than the total energy $E$ released by exothermic reactions.
The typical energy per unit mass released by exothermic reactions was derived from laboratory experiments to be $E/m = 11200$--$18300~\mathrm{kJ~kg^{-1}}$ for irradiated solid methane, $E/m \ga 1500~\mathrm{kJ~kg^{-1}}$ for UV photolyzed ices, $E/m = 3200$--$4000~\mathrm{kJ~kg^{-1}}$ for frozen carbon molecules and noble-gas atoms, and $E/m = 6400$--$8200~\mathrm{kJ~kg^{-1}}$ for amorphous carbon \citep{carpenter1987,shabalin1997,schutte-greenberg1991,wakabayashi-et-al2004,yamaguchi-wakabayashi2004,tanaka-et-al2010}.
Hereafter, we consider the total energy per unit mass of organic refractory material released by exothermic reactions in the range of $E/m \ge 1500~\mathrm{kJ~kg^{-1}}$.

Although we cannot specify the triggering temperature $T_\mathrm{trig}$ without details on reaction kinetics, it should be an equilibrium temperature at a heliocentric distance that is larger than the Saturnian orbit around the Sun.
The equilibrium temperature of a dust particle is calculated by 
\begin{eqnarray}
\Gamma_\mathrm{abs} = \Gamma_\mathrm{rad} + \Gamma_\mathrm{sub}, \label{energy-balance}
\end{eqnarray}
where $\Gamma_\mathrm{abs}$, $\Gamma_\mathrm{rad}$, and $\Gamma_\mathrm{sub}$ denote the heating rate of a dust particle by solar radiation, the cooling rate of a dust particle by thermal radiation, and the cooling rate of a dust particle by sublimation, respectively \citep[see, e.g.,][]{mukai-mukai1973,lamy1974}.

The heating rate of a dust particle by solar radiation is given by
\begin{eqnarray}
\Gamma_\mathrm{abs} = \pi {\left({\frac{R_\sun}{r}}\right)}^{2} \int_{0}^{\infty} C_\mathrm{abs}\left({m^\ast, \lambda}\right)  B_\sun\left({\lambda}\right) \,d\lambda ,
\end{eqnarray}
where $C_\mathrm{abs}\left({m^\ast, \lambda}\right)$ is the cross section of absorption at a wavelength of $\lambda$ and the complex refractive index of $m^\ast$, $B_\sun\left({\lambda}\right)$ is the solar radiance, and $r$ is the heliocentric distance \citep{kimura-mann1998,kimura-et-al2002}.
Computations of $C_\mathrm{abs}\left({m^\ast, \lambda}\right)$ for a silicate core, organic mantle particle were performed in the framework of the Mie theory \citep{aden-kerker1951,bohren-huffman1983}.
Complex refractive indices $m^\ast$ of organic material and of the amorphous silicate necessary for the computations are taken from \citet{li-greenberg1997} and \citet{scott-duley1996}, respectively.
The cooling rate of a dust particle by thermal radiation is given by
\begin{eqnarray}
\Gamma_\mathrm{rad} = 4 \pi \int_{0}^{\infty} C_\mathrm{abs}\left({m^\ast, \lambda}\right)  B\left({\lambda, T}\right) \,d\lambda, \label{radiative-cooling}
\end{eqnarray}
where $B\left({\lambda, T}\right)$ is the Planck function at a temperature of $T$ \citep{kimura-et-al1997,kimura-et-al2002,li-greenberg1998}.
The cooling rate of a dust particle by sublimation is given by
\begin{eqnarray}
\Gamma_\mathrm{sub} = S\,\sqrt{\frac{M_\mathrm{or} u}{2\pi k_\mathrm{B}T}}\,p\left({T}\right) L, \label{sublimation-cooling}
\end{eqnarray}
where $k_\mathrm{B}$ and $u$ are the Boltzmann constant and the atomic mass unit, $S$ is the surface area of the particle, and $L$ and $M_\mathrm{or}$ are the latent heat of sublimation and the molecular weight of organic materials \citep{kimura-et-al1997,kimura-et-al2002,kobayashi-et-al2009}.
The vapor pressure $p\left({T}\right)$ is described by the Clausius-Clapeyron relation as
\begin{eqnarray}
p\left({T}\right) = \exp\left({-\frac{M_\mathrm{or} u}{k_\mathrm{B}T}\,L+b}\right),
\end{eqnarray}
where $b = \ln p\left({\infty}\right)$ is a constant.
Here we represent the thermodynamic properties of interstellar organic materials by those of hexamethylenetetramine (HTM), which is an organic refractory residue from interstellar ice analogs \citep{briani-et-al2013}.
Accordingly, we insert $M_\mathrm{or} = 140$, $L = 5.62 \times {10}^{5}~\mathrm{J\,{kg}^{-1}}$, and $e^{b}=4.24 \times{10}^{12}~\mathrm{Pa}$ into Eq.~(\ref{sublimation-cooling}).

\section{Results}

  \begin{figure}
   \centering
   \includegraphics[width=\hsize]{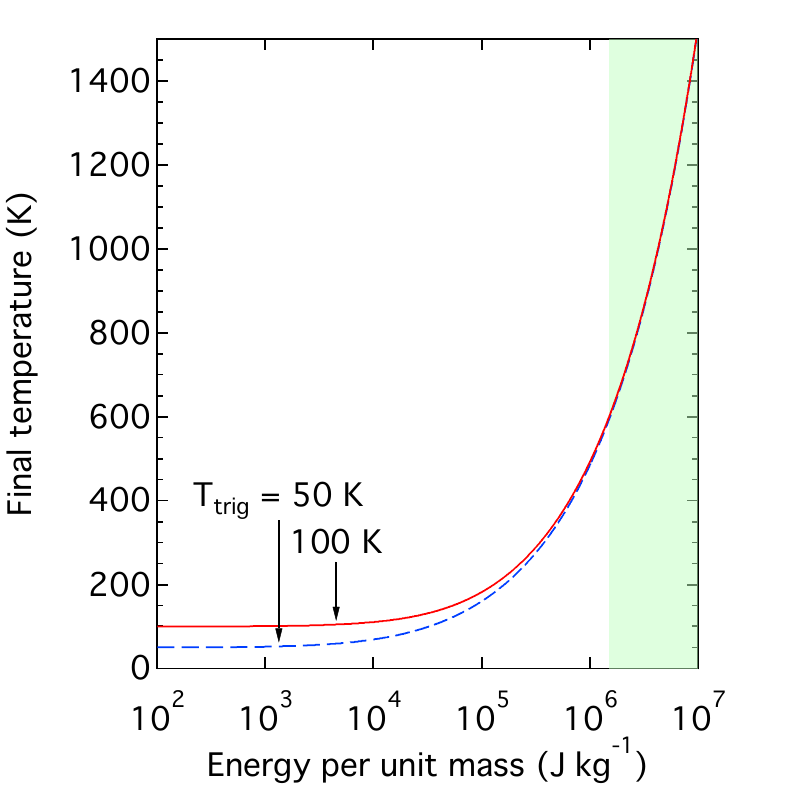}
      \caption{Final temperature of a grain raised by the release of an energy $\varepsilon\left({T}\right)$ per unit mass of organic matter. Solid curve: the triggering temperature of 100~K. Dashed curve: the triggering temperature of 50~K. The shaded area indicates the range of energy $E$ released from a unit mass of organic matter by exothermic chain reactions.
              }
         \label{fig1}
   \end{figure}

Figure~\ref{fig1} shows how the energy required to heat up a unit mass of organic matter to the final temperature, $\varepsilon\left({T}\right)/m$, changes with the temperature $T$ (see Eq.~(\ref{required-energy})).
Since we cannot specify at which temperature exothermic reactions are triggered, we plotted the results with the required energy for two triggering temperatures: $T_\mathrm{trig} = 50~\mathrm{K}$ (dashed line) and $100~\mathrm{K}$ (solid line).
Also, the range of energies released by exothermic
reactions through the recombination of free radicals or the rearrangement
of carbon bonds is enclosed by a shaded area.
If the temperature of sublimation is higher than approximately $200~\mathrm{K}$, then the energy required to raise a dust particle to the temperature of sublimation is almost independent of the triggering temperature.
The total energy released by exothermic reactions is sufficient to heat a dust particle up to the temperature of $T \ga 600~\mathrm{K,}$ regardless of the triggering temperature.

   \begin{figure}
   \centering
   \includegraphics[width=\hsize]{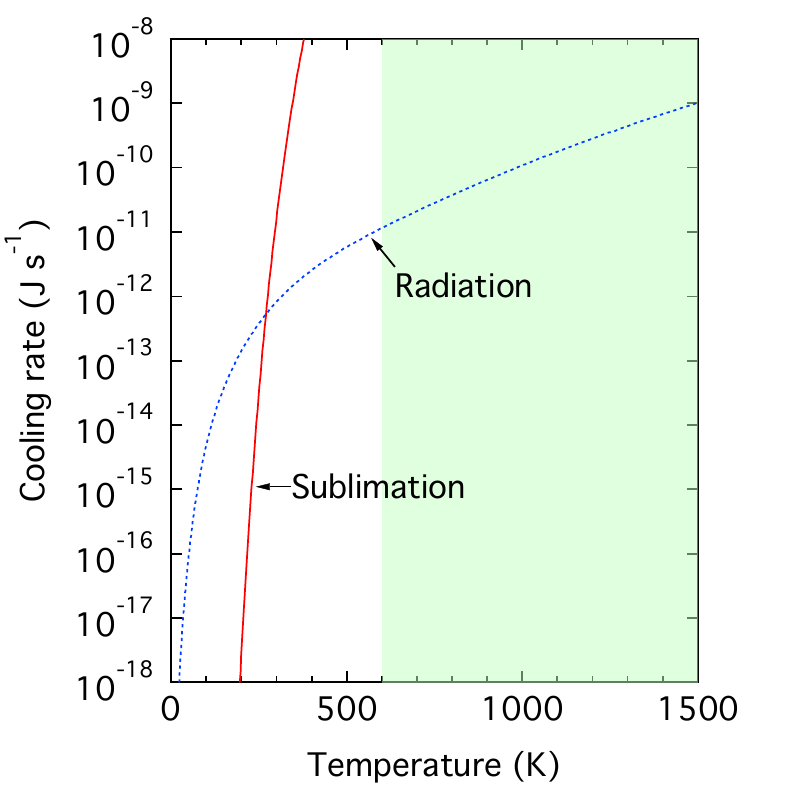}
      \caption{Cooling rates for a grain with a radius of $a=0.1~\mu$m as a function of the grain temperature $T$. Solid curve: sublimation cooling rate $\Gamma_\mathrm{sub}$. Dashed curve: radiative cooling rate $\Gamma_\mathrm{rad}$. The shaded area indicates the range of the final temperature attained by the released energy of $E/m \ge 1500~\mathrm{kJ~kg^{-1}}$ (see Fig.~\ref{fig1}).
              }
         \label{fig2}
   \end{figure}

In the range of temperatures where sublimation dominates the cooling process, $\Gamma_\mathrm{sub}$ should exceed the cooling rate by radiation, $\Gamma_\mathrm{rad}$ \citep{leger-et-al1985}.
Figure~\ref{fig2} shows the sublimation cooling rate $\Gamma_\mathrm{sub}$ and the radiative cooling rate $\Gamma_\mathrm{rad}$ for grains with $a=0.1~\mathrm{\mu m}$ as a function of temperature $T$ (see Eqs.~(\ref{radiative-cooling})--(\ref{sublimation-cooling})).
Our results show that sublimation dominates over radiation for cooling the grains when the temperature of the grains exceeds $T=270~\mathrm{K}$.

   \begin{figure}
   \centering
   \includegraphics[width=\hsize]{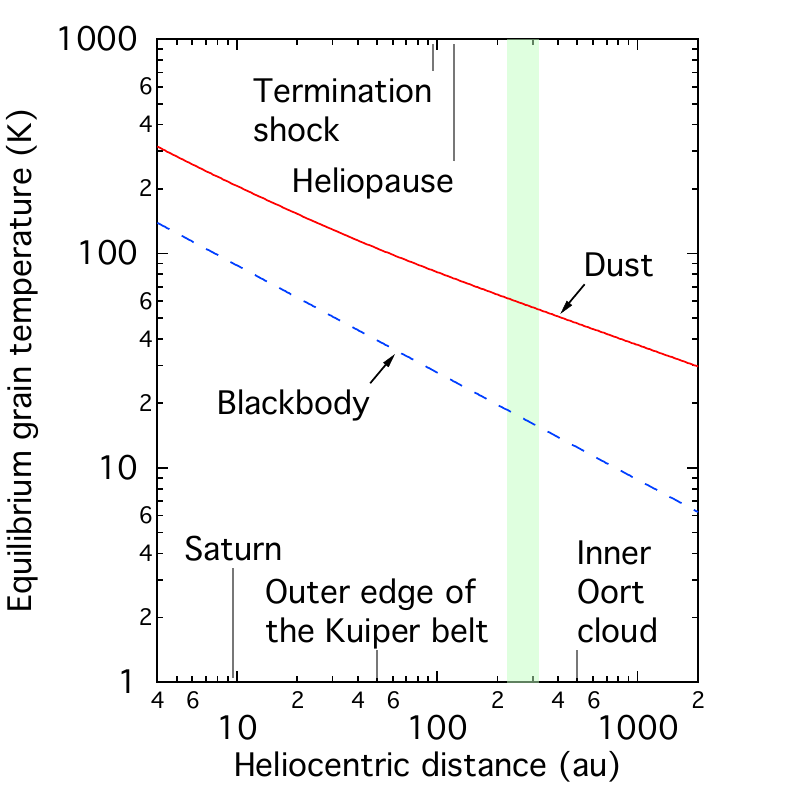}
      \caption{Equilibrium temperature of a grain with a radius of $0.1~\mathrm{\mu m}$ as a function of heliocentric distance (solid line). The dashed line indicates the blackbody temperature. The shaded area indicates the filtration region where a portion of neutral atoms is converted to ions by the charge exchange with ions.
              }
         \label{fig3}
   \end{figure}

In Fig.~\ref{fig3}, we plotted the equilibrium temperature of a dust particle that has $a = 0.1~\mathrm{\mu m}$ as a function of heliocentric distance, computed by Eq.~(\ref{energy-balance}) in the framework of the Mie theory (solid line).
Also, the blackbody temperature of an airless body that is illuminated by the Sun for comparison is plotted as a dashed line.
The filtration region, where a portion of neutral atoms is converted to ions by a charge exchange with ions, is enclosed by a shaded area.
The temperature of the particle beyond Saturn is kept below 200~K, while the blackbody temperature is approximately 90~K in the Saturnian orbit.
The particle attains a temperature of 106~K at 50~AU from the Sun, which is the outer edge of the Kuiper belt, and 47~K at 500~AU, which corresponds to the orbit of inner Oort cloud objects (sednoids).
The equilibrium temperature of the particle at the heliospheric boundary is 77~K at 120~AU and 83~K at 95~AU, which are approximately the heliocentric distances to the heliopause and the termination shock, respectively, in the upwind direction.

\section{Discussion}

\subsection{Sublimation of organic materials}

We have investigated the possibility that organic forming elements desorb completely by sublimation as LIC dust approaches the Sun.
Our results suggest the sublimation of organic refractory material by exothermic chemical reactions with a final temperature of $T \ga 600~\mathrm{K}$.
A comparison of Fig.~\ref{fig1} with Fig.~\ref{fig2} indicates that a dust particle is heated up to the temperature of $T > 270~\mathrm{K}$ if exothermic reactions release the energy of $E/m > 300~\mathrm{kJ~kg^{-1}}$ for $T_\mathrm{trig} = 50~\mathrm{K}$ and $E/m > 300~\mathrm{kJ~kg^{-1}}$ for $T_\mathrm{trig} = 100~\mathrm{K}$.
Since the rearrangement of carbon bonds in amorphous carbon is accompanied with graphitization, this might not be relevant for the desorption of organic forming elements.
The concentration of frozen free radicals are expected in organic forming atoms and molecules ($1$--$10$\%), because of ultraviolet photolysis in the interstellar medium \citep{greenberg1976,sorrell2001}.
Even if the concentration of free radicals is too small to completely sublime all the organic refractory component, built-up pressure by sublimation may induce the desorption of molecules by an explosion \citep{schutte-greenberg1991}.
It is, therefore, not extraordinary that organic refractory material stores the energy of $E/m \ga 270~\mathrm{kJ~kg^{-1}}$, releases it instantaneously by exothermic reactions and shattering, and ends up with its complete desorption.

\subsection{Gas-to-dust mass ratio}

It is worth noting that a signature of organic sublimation in the Solar System may be found as an increase in the gas-to-dust mass ratio, which is $R_\mathrm{g/d} \simeq 121$ in the LIC \citep{kimura2015}.
In fact, \citet{krueger-et-al2015} have derived the gas-to-dust mass ratio of $R_\mathrm{g/d} = 193^{+85}_{-57}$ from the entire data set of Ulysses LIC dust impacts measured within 5~AU from the Sun.
This high mass ratio of LIC gas-to-dust is associated with a low mass density of LIC dust measured in situ by the impact ionization dust detector onboard Ulysses.
According to our model, Ulysses should have detected LIC dust particles that have experienced the sublimation of organic materials at heliocentric distances beyond 10~AU from the Sun.
Therefore, one has to take a weight loss of organic materials from LIC dust into account when estimating the gas-to-dust mass ratio in the LIC from impacts of interstellar dust measured in situ inside the Solar System.
We expect that the sublimation of organic materials elevates the gas-to-dust mass ratio from $R_\mathrm{g/d}  = 121 \pm 22$ to $199 \pm 16$ in the inner Solar System (see Appendix~\ref{appendix_b}).
   \begin{figure}
   \centering
   \includegraphics[width=\hsize]{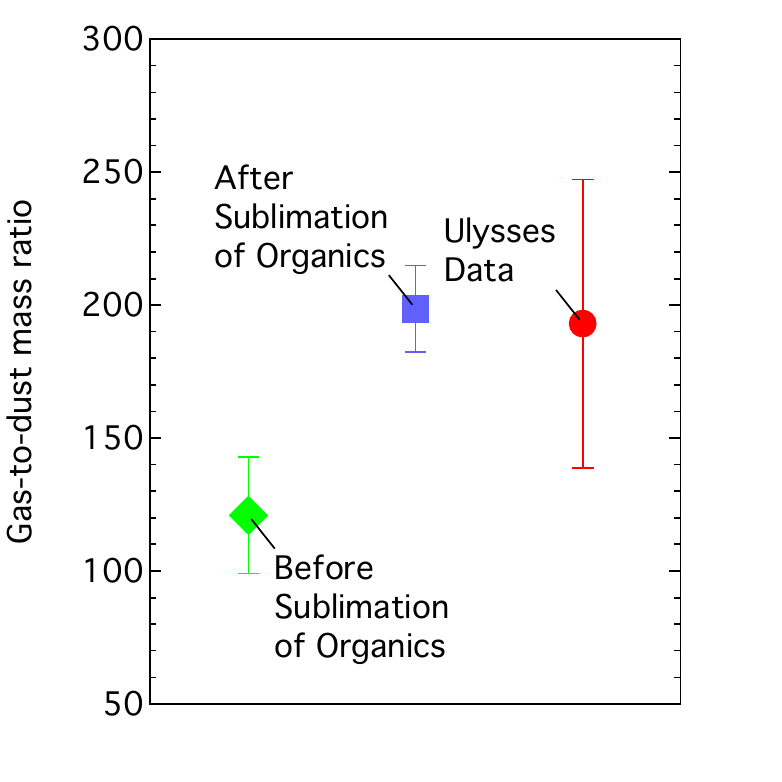}
      \caption{Gas-to-dust mass ratio of the local interstellar cloud (LIC). 
      The filled diamond and square indicate the ratios determined by the depletion of elements in the LIC, before and after the sublimation of organic refractory material (see Appendix~\ref{appendix_b}). 
      The filled circle is the the gas-to-dust mass ratio of the LIC derived from the Ulysses in-situ measurements of LIC dust impacts by \citet{krueger-et-al2015}.
      The error bars represent the standard deviations.
              }
         \label{fig4}
   \end{figure}
The gas-to-dust mass ratio of $R_\mathrm{g/d} \approx 200$ is entirely consistent with the value derived from the Ulysses in-situ measurements of LIC dust impacts.
Although the Ulysses data have large uncertainties, the difference between the Ulysses data and the depletion data after the subtraction of the organic component is not statistically significant at the 5\% level.
Consequently, our results explain the reason that organic forming elements have never been identified in the Stardust samples and the Cassini data of LIC dust, yet they solve the puzzle as to why the mass density of LIC dust is so low in the Ulysses data.

\subsection{Interstellar pickup ions}

   \begin{figure}
   \centering
   \includegraphics[width=\hsize]{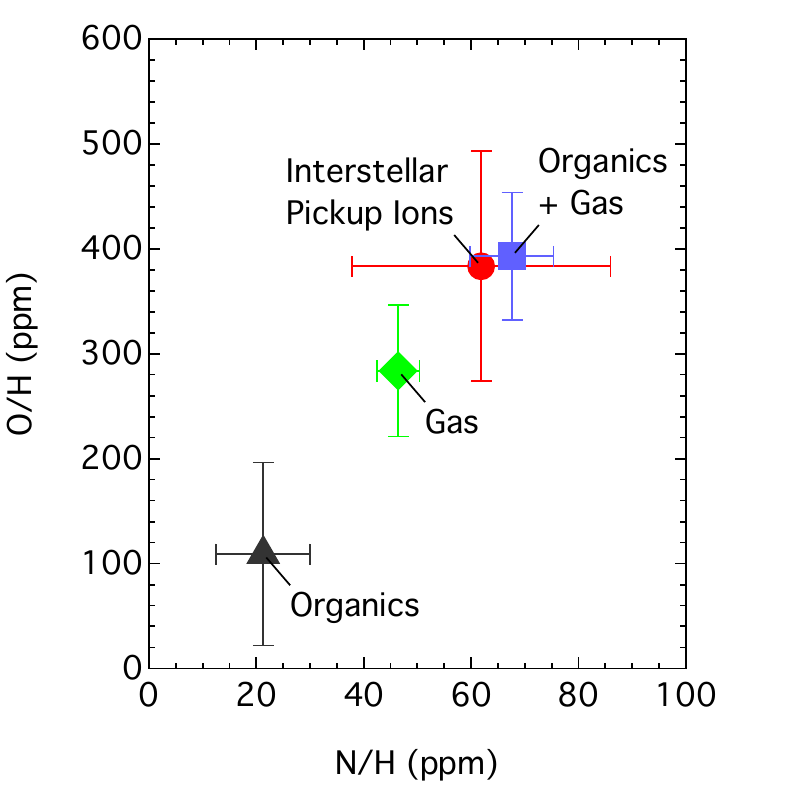}
      \caption{Elemental abundances of nitrogen and oxygen atoms per million hydrogen atoms. 
      The filled triangle, diamond, and square indicate the abundances in the organic material, the gas phase, and the sum of the organic and gas phases, respectively (see Appendix~\ref{appendix_b}). 
      The filled circle is the elemental abundance determined by the measurements of interstellar pickup ions \citep{gloeckler-geiss2004}.
      The error bars represent the standard deviations.
              }
         \label{fig5}
   \end{figure}
So-called interstellar pickup ions detected inside the heliosphere have been used as a powerful tool to study the elemental abundances of gas in the LIC \citep{gloeckler-geiss2004}.
The pickup ion measurements suggest that all nitrogen atoms and most oxygen atoms of the LIC are in the gas phase, contrary to the HST measurements of gas absorption lines.
The sublimation of organic materials could influence the elemental abundances of C, N, O, and H for interstellar pickup ions, since the C, N, O, and H atoms that desorbed from dust particles are also picked up by the solar wind, in the same way as interstellar neutral atoms.
The abundances of elements in the dust phase of the LIC allow us to place significant constraints on the chemical composition of the organic material, although the exact nature of the organic material is unknown (see Appendix~\ref{appendix_a}).
In Fig.~\ref{fig5}, we compare the elemental abundances of nitrogen and oxygen atoms in the organic material (filled square), the gas phase (filled triangle), the sum of organic and gas phases (filled square), and interstellar pickup ions (filled circle).
It turns out that although interstellar pickup ions have an excess of the N and O, in comparison to LIC gas, each pickup-ion abundance is remarkably closer to the sum of organic and gas phase abundances, compared with the gas-phase abundance alone.
We find that the sum of organic and gas phase abundances is not statistically different at the 5\% level of significance from both N and O abundances of interstellar pickup ions.
Although the large uncertainties in the measured N/H ratio of interstellar pickup ions conceal the deviation from LIC gas, the O/H ratio of interstellar pickup ions differs from that of LIC gas at the 95\% confidence level. 
Consequently, the sublimation of organic forming elements in the Solar System would be in good harmony with the elemental abundances of interstellar pickup ions measured inside the Solar System.

\subsection{Triggering temperature}

The process behind sublimation is that the heat of the Sun triggers exothermic chain reactions by the recombination of free radicals or the rearrangement of carbon bonds.
Because the triggering temperature depends on the organic composition that is unknown for LIC dust, it is difficult, if not impossible, to determine the triggering temperature precisely.
\citet{schutte-greenberg1991} estimated the triggering temperature to lie in the range of $T_\mathrm{trig} = 24.5$--28~K for free radicals in their UV-photolyzed ice mixtures, similar to $T_\mathrm{trig} = 27~\mathrm{K}$, which was measured by \citet{dhendecourt-et-al1982} with different UV-photolyzed ice mixtures. 
Since organic forming elements are strongly depleted from LIC dust, which is already at the orbit of Saturn according to the Cassini in-situ measurements, exothermic reactions should have been triggered beyond 10~AU from the Sun.
However, the N and O abundances of interstellar pickup ions may not be well accounted for by the sum of organic materials and gas, as shown in Fig.~\ref{fig4}, unless exothermic reactions are triggered before crossing the so-called filtration region.
The filtration region extends to a circum-heliospheric interstellar medium of 100--200~AU beyond the heliopause and converts a portion of neutral atoms to ions by a charge exchange with ions \citep{gloeckler-et-al1997,izmodenov-et-al2004}.
This implies that the triggering temperature should be lower than at least $T = 77~\mathrm{K}$, which is the equilibrium temperature at the heliopause (see Fig.~\ref{fig3}).
It is, however, reasonable to assume that the triggering temperature lies below the equilibrium temperature just beyond the filtration region, which is close to the inner edge of the Oort cloud.
Therefore, we suggest $T_\mathrm{trig} \la 50~\mathrm{K}$, because it does not contradict the triggering temperatures for the recombination of free radicals and the rearrangement of carbon bonds.
Since the temperature of interstellar dust is kept as low as 18~K in the LIC, we propose that the triggering temperature lies in the range of $T_\mathrm{trig} = 20$--50~K.

\subsection{Annealing of amorphous silicates}

   \begin{figure}
   \centering
   \includegraphics[width=\hsize]{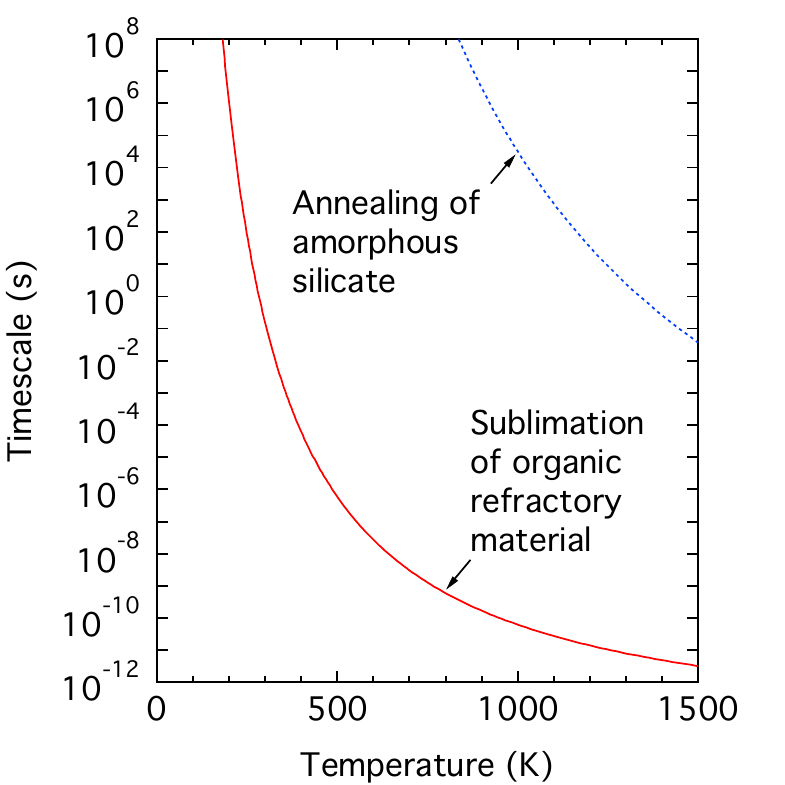}
      \caption{Timescales $\tau_\mathrm{c}$ for annealing of amorphous silicate (solid curve) and those $\tau_\mathrm{sub}$ for the sublimation of organic refractory material (dashed curve).
              }
         \label{fig6}
   \end{figure}
While exothermic reactions elevate a dust particle to the sublimation temperature, it is of great importance to find out whether the annealing of amorphous silicates takes place or not.
Experimental and theoretical studies on the crystallization of amorphous silicate grains covered by CH$_4$-doped amorphous carbon have shown that exothermic reactions in the mantle result in the crystallization of the silicate core at room temperature \citep{kaito-et-al2007,tanaka-et-al2010}.
Although there is no evidence for the annealing of amorphous silicates of LIC dust en route to the Solar System, we examine whether the amorphous silicate core of a submicron grain crystallizes by exothermic reactions in its organic mantle.
We have found that the energy released by exothermic reactions is sufficient to heat up a dust particle to the temperature of $T \ga 270~\mathrm{K}$ where sublimation dominates the cooling (see Figs.~\ref{fig1} and \ref{fig2}).
If the annealing of amorphous silicates proceeds with crystallization at $T = 1000~\mathrm{K}$, it takes $\tau_\mathrm{c} = 4.4\times{10}^{3}$--$9.1\times{10}^{4}~\mathrm{s}$, depending on the silicate mineralogy \citep{fabian-et-al2000,kimura-et-al2002,tanaka-et-al2010}.
If we estimate the timescale for sublimation by $\tau_\mathrm{sub} = mxL/\Gamma_\mathrm{sub}$, it turns out that organic materials completely sublimate within $\tau_\mathrm{sub} = 6.3\times{10}^{-11}~\mathrm{s}$ at $T = 1000~\mathrm{K}$.
The duration of organic sublimation is drastically shorter than the timescale for silicate crystallization, even if we consider higher temperatures (see Fig.~\ref{fig6}).
Therefore, we may assert that the annealing of amorphous silicate cores does not take place during the sublimation of organic materials.

\subsection{Large grains of micrometer sizes}

We have restricted our discussion on the sublimation of organic materials to submicron grains, as this is the size regime of LIC dust that is compositionally analyzed by Cassini's Cosmic Dust Analyzer (CDA), but micrometer-sized grains are certainly present in the LIC as detected by Ulysses within 5~AU from the Sun \citep{gruen-et-al1994,krueger-et-al2015}.
If LIC dust particles of micrometer sizes are agglomerates of submicron core-mantle grains, as suggested by \citet{kimura-et-al2003b}, then their temperatures are most likely close to those of submicron core-mantle grains.
Therefore, we expect that the sublimation of organic materials from the constituent grains of the agglomerates takes place in a similar way as we have studied in this work.
Once organic mantles of submicron grains in agglomerates proceed with sublimation, then the so-called packing effect may produce compact agglomerates of silicate core grains \citep{mukai-fechtig1983}.
Therefore, there is a possibility that such compact agglomerates of silicate grains are the micrometer-sized interstellar grains detected by Ulysses within 5~AU from the Sun.
Laboratory analyses of Stardust samples have revealed that micrometer-sized interstellar grains are indeed characteristic of low-density silicate materials, which closely resemble agglomerates of silicate grains \citep{butterworth-et-al2014,westphal-et-al2014a,westphal-et-al2014b}.
The extraordinary low capture velocities of the Stardust interstellar grains and high ratios of solar radiation pressure to solar gravity on these grains found by \citet{postberg-et-al2014} and \citet{sterken-et-al2014} also point to low-density agglomerates of silicate grains with metallic inclusions \citep{kimura2017}.
Alternatively, if the micrometer-sized interstellar grains are single micrometer-sized particles with an organic mantle and a silicate core, then their temperatures are lower than submicron core-mantle grains.
This indicates that the organic component of micrometer-sized interstellar grains sublimates at smaller heliocentric distances than the distances estimated from Fig.~\ref{fig3}.
Nevertheless, we are confident that organic materials sublimate beyond 10~AU since even micrometer-sized grains are still much warmer than the blackbody.
Consequently, we conclude that micrometer-sized grains that are registered on the dust detector onboard Ulysses do not retain organic compounds either.

\subsection{Physical and chemical structures of interstellar dust}

Although we assume that interstellar dust consists of an amorphous silicate in the core of the particle and an organic material in the mantle, the physical structure of interstellar dust and the chemical structure of carbonaceous matter are open to debate.
\citet{mathis-whiffen1989} proposed interstellar dust to be conglomerates of silicate grains, graphite grains, and amorphous carbon grains because graphite grains and amorphous carbon grains are condensed in carbon-rich envelopes around evolved stars.
While exothermic reactions may take place in amorphous carbon grains by rearrangements of carbon atoms, the outcome would be the phase-transition of amorphous carbon to graphite \citep{wakabayashi-et-al2004}.
If conglomerates of silicate grains, graphite grains, and amorphous carbon grains represent realistic structures of interstellar dust, the presence of carbon atoms from graphite in such a high abundance should have been identified in the data of Helios, Cassini, and Stardust.
Since this is not the case, we could rule out that conglomerates of silicate grains, graphite grains, and amorphous carbon grains are the major population of dust in the LIC.
We would like to point out that the depletions of C, N, and O in the gas phase of the LIC are not in harmony with pure carbon, such as graphite and amorphous carbon, but with organic refractory materials \citep{kimura-et-al2003b,kimura2015}.
The reason that we adopt the core-mantle structure is that it is the outcome of ice accretion in molecular clouds and subsequent photo-processing in the DISM, while the growth of core-mantle grains into agglomerates leaves room for discussion \citep{li-greenberg1997,kimura-et-al2003b,kimura2017}.
Unfortunately, we cannot ensure the uniqueness of the structure, but the desorption of organic-forming elements from the grain surface is a novel idea to solving a number of conundrums and thus the model of core-mantle grains provides a feasible solution to the observations of missing organic materials in interstellar dust penetrating into the Solar System.
Because the available body of facts do not conflict with the picture that LIC dust is composed of organic refractory material in its mantle and amorphous silicate in its core, the sublimation of organic matter sheds light on the substantial depletion of organic forming elements in LIC dust observed inside the heliosphere.

\subsection{Delivery of interstellar organic matter to the early Earth}

The delivery of extraterrestrial organic compounds to the primitive Earth might have played a crucial role in the origin of life, on the assumption that they are the seeds for prebiotic chemistry \citep{anders1989,chyba-et-al1990}.
Comets and asteroids, indeed, transport extraterrestrial organic matter to the Earth by means of interplanetary dust particles and chondritic meteorites, but the transportation of extraterrestrial seeds to the Earth from the interstellar medium is an unresolved issue.
In contrast to extraterrestrial abiotic organic seeds, panspermia is the hypothesis that attributes the origin of life on Earth to the delivery of biotic material, namely, microorganisms via interplanetary and/or interstellar space \citep{arrhenius1908}.
We have shown that organic compounds of interstellar dust cannot remain intact, but they suffer from sublimation prior to its entry into the heliosphere because solar heating facilitates exothermic chain reactions of radicals.
Contrary to the interstellar organic matter, comets retain refractory organic matter without a threat to sublimation in a near-Earth orbit since the region of comet formation in the solar nebula was warm enough to exhaust radicals and hence prevent their accumulation \citep{barnun-kleinfeld1989}. 
Therefore, unlike comets and asteroids, the delivery of extraterrestrial organic matter to the Earth via the DISM is very difficult, if not impossible, irrespective of its biotic or abiotic nature.
If one of the future exploration missions would find evidence for the sublimation of interstellar organic matter as indicated by our results, the hypothesis of panspermia from the DISM must be discarded.

\begin{acknowledgements}
Special thanks are due to Harald Kr\"{u}ger who provided stimulating discussions and Ulysses results on LIC dust impacts prior to publication of the results.
We are also grateful to an anonymous reviewer for her/his fruitful comments that helped us to improve the manuscript. 
H. Kimura thanks Martin Hilchenbach and Thomas Albin for their hospitality during his stay at Max Planck Institute for Solar System Research (MPS), where much of the writing was performed, as well as Hitoshi Miura and Andrew Westphal for their useful correspondences.
The author also thanks MPS's research fellowship and JSPS's Grants-in-Aid for Scientific Research (\#26400230, \#15K05273, \#19H05085).
\end{acknowledgements}

%-------------------------------------------------------------------

\begin{appendix} %First online appendix
\section{The composition of organic refractory material in the local interstellar cloud (LIC)\label{appendix_a}}

The arguments of missing atoms allowed us to derive the abundances of elements in the dust phase of the LIC from the depletion of the elements in the gas phase of the cloud \citep{spitzer1978}.
We determine the composition of organic refractory material in the LIC based on the assumption that the organic refractory material is the main carrier of the missing O atoms as well as C and N atoms \citep[see][]{whittet2010}.
As the main carriers of Mg, Al, Si, S, and Fe, we consider MgAl$_2$O$_4$, FeNi, Mg$_2$SiO$_4$, MgSiO$_3$, and FeS, which are consistent with the returned samples of LIC grains \citep{westphal-et-al2014b}. 
The distribution of the elements in their main dust-phase carriers can be determined uniquely, once the depletion of the elements from the gas phase is given \citep{kimura2015}.
Table~\ref{tablea1} gives the resulting distribution of elements in their main dust-phase carriers as well as the composition of organic refractory material, denoted by CHON.
%
%-------------------------------------------------------------
%                                             Two column Table 
%-------------------------------------------------------------
%
\begin{table*}
 \caption[]{\label{tablea1}Abundances of the elements per million hydrogen atoms and their plausible main carriers in the dust phase of the local interstellar cloud (LIC).}
\begin{tabular}{lccccccccc}
 \hline \hline
\noalign{\smallskip}
 Element & \multicolumn{6}{c}{Compound} & Dust & Gas & Sun \\
\noalign{\smallskip}
  & MgAl$_2$O$_4$ & FeNi & Mg$_2$SiO$_4$ & MgSiO$_3$ & FeS & CHON & & & 
 \\ \hline
C   & $0$ & $0$ & $0$ & $0$ & $0$ & $100.89^{+51.76}_{-51.76}$ & $100.89^{+51.76}_{-51.76}$ & $168.27^{+41.46}_{-41.46}$ & $269.15^{+32.84}_{-29.27}$\\
N   & $0$ & $0$ & $0$ & $0$ & $0$ & $21.24^{+8.73}_{-8.73}$ & $21.24^{+8.73}_{-8.73}$ & $46.37^{+3.96}_{-3.96}$ & $67.61^{+8.25}_{-7.35}$\\
O   & $5.42^{+0.39}_{-0.39}$ & 0 & $32.55^{+17.36}_{-17.36}$ & $58.57^{+14.73}_{-14.73}$ & $0$ & $109.39^{+87.33}_{-87.33}$ & $205.92^{+84.31}_{-84.31}$ & $283.86^{+62.68}_{-62.68}$ & $489.78^{+59.76}_{-53.26}$\\
Mg   & $1.35^{+0.10}_{-0.10}$ & $0$ & $16.27^{+8.68}_{-8.68}$ & $19.52^{+4.91}_{-4.91}$ & $0$ & $0$ & $37.15^{+3.68}_{-3.68}$ & $2.66^{+0.33}_{-0.33}$ & $39.81^{+3.84}_{-3.50}$\\
Al   & $2.71^{+0.20}_{-0.20}$ & $0$ & $0$ & $0$ & $0$ & $0$ & $2.71^{+0.20}_{-0.20}$ & $0.11^{+0.03}_{-0.03}$ & $2.82^{+0.20}_{-0.19}$\\
Si   & $0$ & $0$ & $8.14^{+4.34}_{-4.34}$ & $19.52^{+4.91}_{-4.91}$ & $0$ & $0$ & $27.66^{+2.30}_{-2.30}$ & $4.70^{+0.52}_{-0.52}$ & $32.36^{+2.31}_{-2.16}$\\
S   & $0$ & $0$ & $0$ & $0$ & $2.59^{+1.42}_{-1.42}$ & $0$ & $2.59^{+1.42}_{-1.42}$ & $10.59^{+1.09}_{-1.09}$ & $13.18^{+0.94}_{-0.88}$\\
Fe   & $0$ & $27.72^{+3.24}_{-3.24}$ & $0$ & $0$ & $2.59^{+1.42}_{-1.42}$ & $0$ & $30.31^{+2.91}_{-2.91}$ & $1.31^{+0.08}_{-0.08}$ & $31.62^{+3.05}_{-2.78}$\\
\hline
\end{tabular}
\tablefoot{The solar and gas-phase abundances of elements are taken from \citet{asplund-et-al2009} and \citet{kimura-et-al2003a}, respectively.}
\end{table*}
Comparisons between mid-infrared spectra of the diffuse interstellar medium (DISM) and of analog organic materials suggest that, on average, the organic refractory component of the DISM is poor in nitrogen and oxygen \citep{pendleton-allamandola2002}.
Because Table~\ref{tablea1} shows that the organic refractory material in the LIC is rich in nitrogen and oxygen, the organic refractory component of the LIC does not appear to have an average elemental composition in the DISM.
However, it should be noted that the composition of the DISM has traditionally been derived from long sightlines\footnote{For example, \citet{pendleton-allamandola2002} used the spectra of the interstellar medium toward the distant luminous star Cyg OB2 No. 12 located at a distance of $1.7 \pm 0.2~\mathrm{kpc}$.} through not only diffuse clouds but also molecular clouds and highly ionized $\mathrm{H}_2$ regions where the dust is, to a great extent, destroyed.
Therefore, the composition of the classical DISM cannot be free of any influence from both molecular clouds and $\mathrm{H}_2$ regions, while the LIC of a few parsecs in scale provides us with an opportunity to study the composition of solely diffuse clouds.
This does not necessarily guarantee the composition of the LIC to be typical for diffuse clouds, but the organic refractory component of the LIC does not seem peculiar at all, as it compositionally resembles the organic refractory material of cometary dust well \citep[see][]{kimura-et-al2003b}. 

\section{The gas-to-dust mass ratio in the local interstellar cloud (LIC)\label{appendix_b}}

The gas-to-dust mass ratio $R_\mathrm{g/d}$ of the LIC can be derived from the eighth column of Table~\ref{tablea1}, which results in $R_\mathrm{g/d} = 120.96\pm{21.83}$.
If the sublimation of organic refractory material takes place, then all the C, N, and O atoms in the organic refractory material are returned to the gas phase.
As a result, the sublimation raises the gas-phase abundances of C, N, and O to $\mathrm{(C/H)_{gas}} = (269.15\pm{30.99})\times{10}^{-6}$, $\mathrm{(N/H)_{gas}} = (67.61\pm{7.78})\times{10}^{-6}$, and $\mathrm{(O/H)_{gas}} = (393.24\pm{60.81})\times{10}^{-6}$, respectively, and the gas-to-dust mass ratio to $R_\mathrm{g/d} = 198.68\pm{16.31}$.

\end{appendix}

\end{document}